# Prediction of new multiferroic and magnetoelectric material $Fe_3Se_4$

Deobrat Singh[1], Sanjeev K. Gupta[2,*], Haiying He[3] and Yogesh Sonvane[1,*]

[1]Advance Materials Lab, Department of Applied Physics, S. V. National Institute of Technology, Surat 395007, India.

[2]Computational Materials and Nanoscience Group, Department of Physics and Electronics, St. Xavier's College, Ahmedabad 380009, India.

[3]Department of Physics and Astronomy, Valparaiso University, Valparaiso, Indiana 46383, USA.

## Abstract

Nowdays, multiferroic materials with magnetoelectric coupling have many real-world applications in the fields of novel memory devices. It is challenging is to create multiferroic materials with strongly coupled ferroelectric and ferrimagnetic orderings at room temperature. The single crystal of ferric selenide ($Fe_3Se_4$) shows type-II multiferroic due to the coexistence of ferroelectric as well as magnetic ordering at room temperature. We have investigated the lattice instability, electronic structure, ferroelectric, ferrimagnetic ordering and transport properties of ferroelectric metal $Fe_3Se_4$. The density of states shows considerable hybridization of Fe-3d and Se-4p states near the Fermi level confirming its metallic behavior. The magnetic moments of Fe cations follow a type-II ferrimagnetic and ferroelectric ordering with a calculated total magnetic moment of 4.25 $\mu_B$ per unit cell ($Fe_6Se_8$). The strong covalent bonding nature of Fe-Se leads to its ferroelectric properties. In addition, the symmetry analysis suggests that tilting of Fe sub-lattice with 3d-$t_{2g}$ orbital ordering is due to the Jahn-Teller (JT) distortion. This study provides further insight in the development of spintronics related technology using multiferroic materials.

*Corresponding authors: sanjeev.gupta@sxca.edu.in (Dr. Sanjeev Gupta)

yas@phy.svnit.ac.in (Dr. Yogesh Sonvane)

**Introduction**

Recent year, multiferroic and magnetoelectric materials are becoming more and more indispensable for many forms of current multi-functional technology, such as highly sensitive magnetic field sensor, filter transducers, filters, phase shifters, memory devices and oscillators[1-4]. For practical applications, we needs to discover multiferroic materials at room temperature which is strongly coupled ferroelectric and ferrimagnetism ordering. In addition to their important polarization properties, ferroelectrics are also pyroelectric[5], where it develops a voltage across the material upon heating, while in the case of piezoelectric a voltage is developed in response to strain across the material. These properties allow ferroelectric materials to be utilized in many device applications, including non-volatile memory[6], thermal detectors[7], piezoelectric applications[8], and energy harvesters[9]. The coexistence of magnetism and ferroelectricity in a material is called multiferroicity, which is of even greater technological and fundamental importance. This has added other potential applications. For instance, it can be utilized in data storage systems, which count both magnetic and electronic states of the compound to store information, and in the magneto-electronic devices using multiferroic thin films[10].

Some transition metal compounds, such as $BiMnO_3$ and $BiFeO_3$ with magnetic $Mn^{3+}$ and $Fe^{3+}$ ions, are ferroelectric[11], owing to the intricate interplay between spin, charge, orbital and lattice degree of freedom in these materials[12]. The mechanisms behind ferroelectricity, however, are not fully understood because of the complicated structures of early ferroelectrics (*i.e.* $BaTiO_3$[13]). The compound $CaMnO_3$ is also an interesting counter example which shows ferroelectricity and magnetic ordering due to the $d^3$ configuration of Mn atoms[14]. It is worthy to be noted that in general the ferroelectricity behavior requires nearly unoccupied subshell orbitals of the transition metal cations, while partially filling d orbitals are required to have magnetic moments. This dilemma largely hinders the coexistence of ferroelectricity and magnetism in a material and explains the scarcity of multiferroic materials[15].

Recently, Bishwas *et al.*,[16] studied the synthesized manganese doped iron selenide nanostructures in an attempt to increase the energy product. Another research group Shao-jie Li *et al.*,[17] obtained high Curie temperature and coercivity performance of $Fe_{3-x}Cr_xSe_4$ nanostructures which can be utilized in the alternative low-cost hard-magnetic materials. Gen Long *et al.*,[18] demonstrated that at low temperatures, $Fe_3Se_4$ nanostructures exhibit giant coercivity. It was proposed that this unusually large coercivity originates from the large magneto-crystalline anisotropy of the monoclinic structure of $Fe_3Se_4$ with ordered Fe vacancies. The ferromagnetic material $BaFe_2Se_3$ is of particular interest because it breaks the parity symmetry of the crystal structure and displays exchange striction effects[19]. Indeed, the iron displacements in the crystal structure are prominent, as revealed by neutron studies[20, 21]. Dong *et al.*[22] showed that the first nearest-neighbor distances between Fe (↑) and Fe (↑) [or Fe (↓) and Fe (↓)] at 0 K become 3.14 Å, much larger than the Fe (↑) and Fe (↓) distance 2.88 Å. On the other hand, this exchange striction is not sufficient to induce Ferroelectric (FE) and Polarization (P) since it breaks parity but not space-inversion symmetry. Most ferroelectric materials are transition metal oxides having vacant d subshells. In this respect the two configurations are not so different, but study of the difference in filling of the d-orbitals as theoretically expected for ferroelectricity and magnetism makes these two ordered states mutually exclusive. According to Giovannetti *et. al.*[24], the ferroelectric and metallic state are found in $LiOsO_3$ material. Interestingly, however, it has been suggested that several other metallic transitions could be "ferroelectric"[23] such as the transition in $LiOsO_3$[24] due to the appearance of a polar axis.

In this work, we focus on the $Fe_3Se_4$ material derived from the monoclinic phase. We have study the lattice instabilities including ferroelectric distortions in monoclinic $Fe_3Se_4$ materials. We have performed first principles calculations within the spin-polarized density functional theory (DFT) framework to obtain the electronic band structure, the role of electron correlations in the metallic state, magnetic properties of the $Fe_3Se_4$ material.

**Methodology**

In this work, all calculations are based on the density functional theory (DFT)[25] as implemented in the Quantum Espresso (QE) package[26]. The Kohn-Sham equations were solved using the Perdew-Burke-Ernzerhof (PBE) exchange-correlation functionals[27] formulated within the generalized gradient approximation (GGA) scheme[28]. We have included sixteen valence electrons for Fe ($3s^2$, $3p^6$, $3d^6$, $4s^2$) and six valence electrons for Se ($4s^2$, $4p^4$) in our calculations. The kinetic energy cut-off is set to 60 Ry which yielded good convergence in results of energies and ground state structural parameters. For structural optimization we have used the conjugate gradient algorithm[29]. The lattice parameters and the internal coordinates of atoms were optimized within the space group of I12/m1 (monoclinic) with the criteria for force and pressure below $10^{-3}$ a.u. A Monkhorst-Pack k-point mesh[30] of 9×9×11 was used for electronic band structure and density of state calculations, while the spontaneous polarization was calculated using the Berry phase method[31] with a k-point mesh of 5×5×7. All the calculations were spin-polarized. Scalar relativistic Troullier-Martins ultra-soft pseudopotentials[32] were employed with non-linear core corrections[33].

**Results and Discussion**

**Structural and electronic properties**

As a benchmark test for the approach used in this work, we have investigated the structural,

electronic and magnetic properties of the monoclinic $Fe_3Se_4$ phase. $Fe_3Se_4$ adopts a normal spinel structure as shown in Fig. 1 and the calculated lattice parameters are listed in Table 1. The $Fe^{3+}$ ions are located in the octahedral sites of the monoclinic lattice formed by $Se^{2-}$ anions forming magnetic ordering. The calculated structural parameters are in very good agreement with previous reported values (Table 1).

**Table 1.** Optimized structural lattice constants of monoclinic $Fe_3Se_4$.

| References | a (Å) | b (Å) | c (Å) | β (°) |
|---|---|---|---|---|
| Present | 6.100 | 3.520 | 11.030 | 91.15 |
| Exp[34,35,16] | 6.208, 6.167, 6.159 | 3.525, 3.537, 3.493 | 11.2832, 11.170, 12.730 | 92, 92, - |

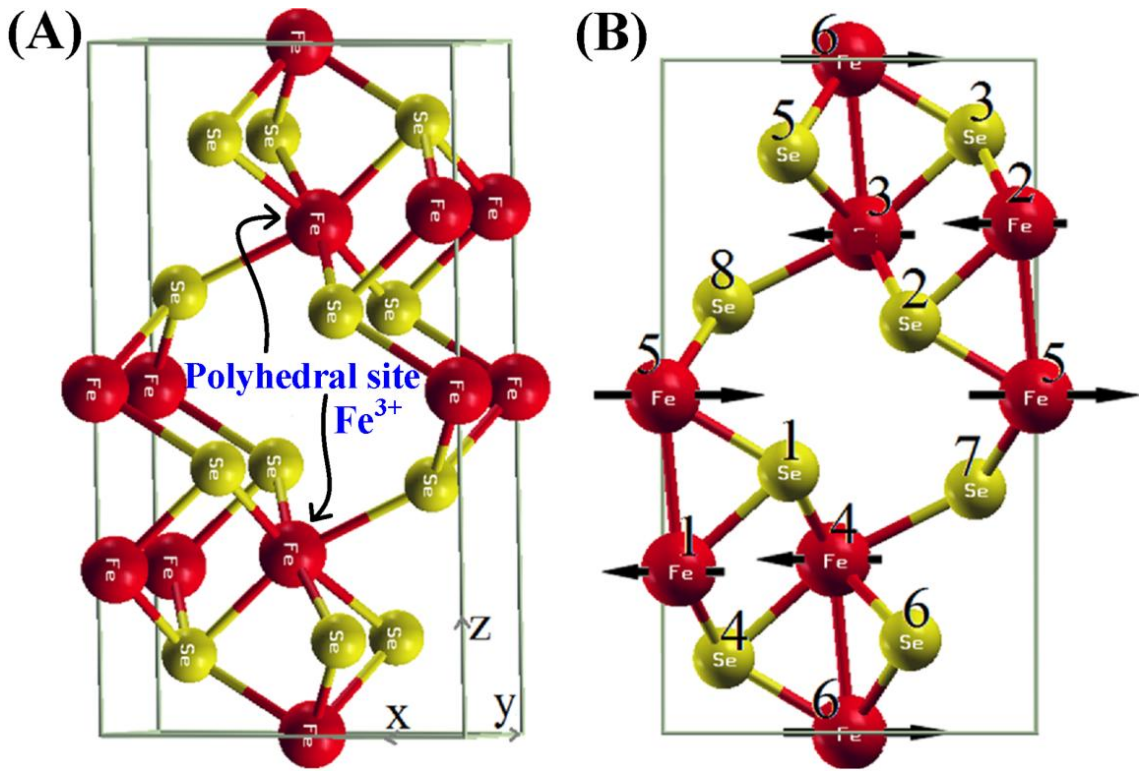

**Figure 1**. (A) Relaxed structure of monoclinic $Fe_3Se_4$ (Fe in red and Se in yellow); (B) distribution of electric dipole moments (black arrows) of Fe atoms due to partial ionic displacement along the *a* direction).

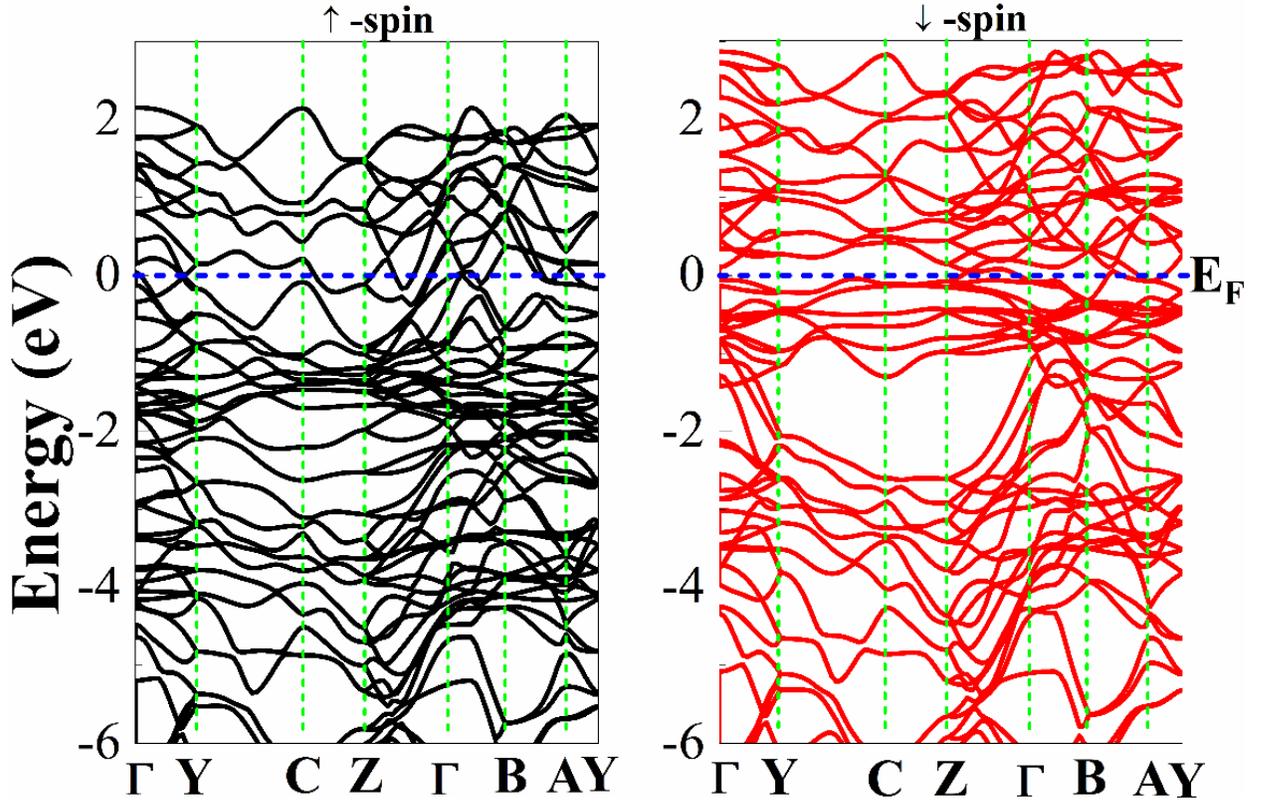

**Figure 2**. The electronic band structure of monoclinic $Fe_3Se_4$ with up spin (left side) and down spin (right side) shown separately.

The electronic band structure of monoclinic $Fe_3Se_4$ is plotted in Figure 2. The Fermi level $E_F$ is set to zero. It is clearly shown that there are bands crossing at $E_F$, thereby the band gap is zero and $Fe_3Se_4$ is metallic in nature. This is further confirmed from the density of states (DOS) plots (see Figure S1 in ESI), which demonstrate asymmetric ↑-spin and ↓-spin DOS. A careful look of the projected density of states (PDOS) reveals ferrimagnetic spin ordering between the iron and selenium atoms. The selenium atoms carried a small magnetic moment (0.14 $\mu_B$/unit cell) due to charge transfer and geometric distortion. Therefore, monoclinic $Fe_3Se_4$ shows metallic behavior in the ferrimagnetic (FM) state.

In order to check whether the metallic behavior is an artifact of neglecting the spin-orbit coupling (SOC), we have further calculated electronic band structure of $Fe_3Se_4$ with the inclusion of SOC interaction (shown in Figure S2 in ESI). The spin-orbit energy splitting is larger at points of high-symmetry. For instance, the spin-orbit splitting at gamma point is about 14 meV and 30 meV in valence band maximum (VBM) and conduction band minimum (CBM), respectively. Since the monoclinic $Fe_3Se_4$ lattice has inversion symmetry, due to the Kramers degeneracy[36], each energy band line is at least doubly degenerate for both spin states. Our results show that the inclusion of the SOC interaction influences only three p-energy bands of Se atoms and five d-energy bands of Fe atoms and the metallic behavior remains unchanged.

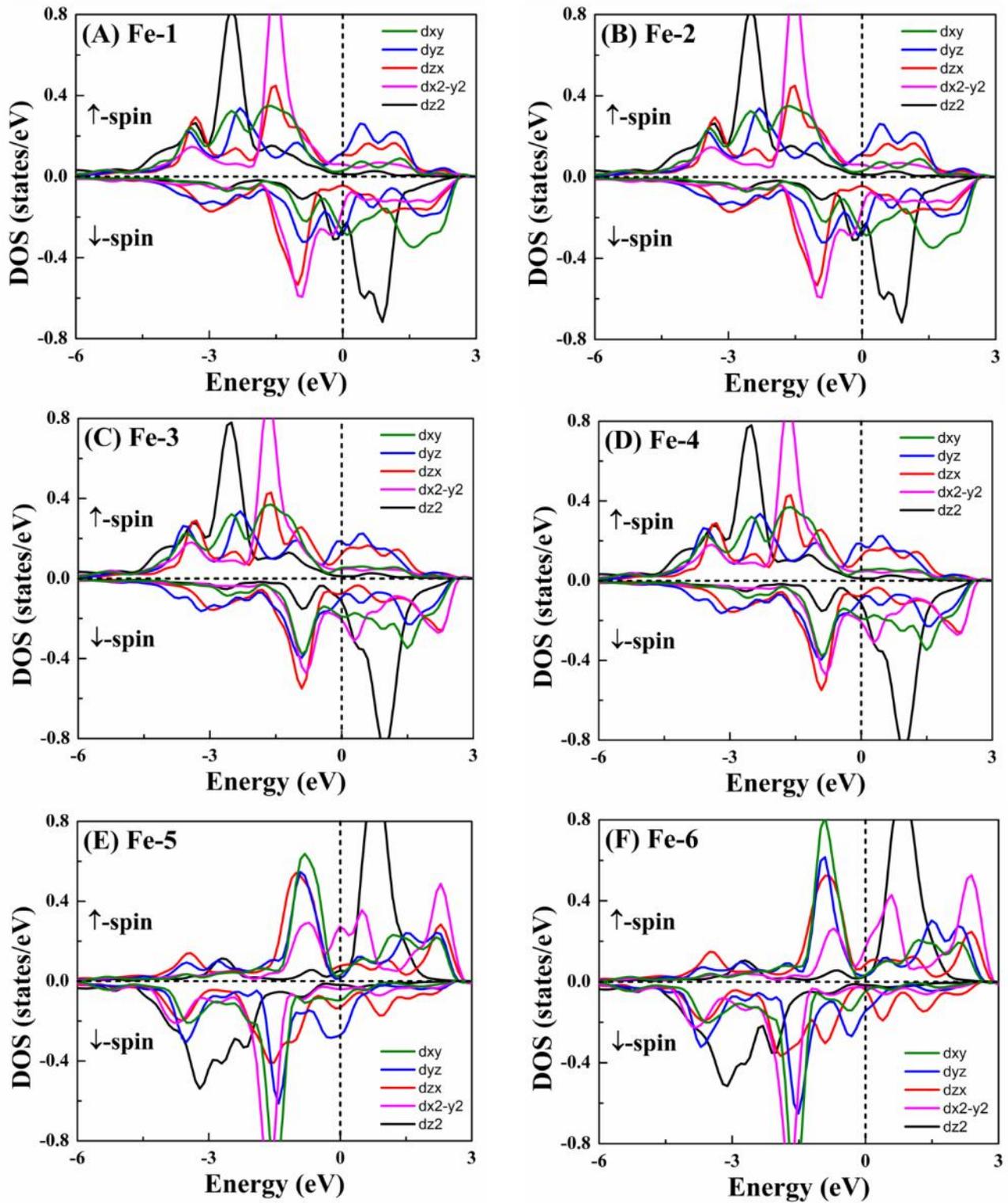

**Figure 3**. PDOS of Fe ions in relaxed $Fe_3Se_4$. The numbering of Fe is the same as in Figure 1(B). Fe-1. Fe-2 and Fe-5 forms one layer, while Fe-3, Fe-4 and Fe-6 forms another.

The origin of the metallic behavior of the monoclinic $Fe_3Se_4$ can also be understood from the PDOS plots (Figure S1 in ESI). It is clear that the spin-up and spin-down states determining the FM metallicity mainly originate from the edge Fe-3d and edge Se-4p orbitals. Fe atoms contribute more to DOS than the Se atoms at Fermi level and the majority of the density of states near the Fermi level for $Fe_3Se_4$ is attributed to the Fe-3d states. The Se-3p bands overlap with the Fe-3d bands in the -7 eV to 3 eV energy range, representing a hybridization of the Se-3p and Fe-3d states to form the covalent bonding while Se-3p and Se-3s orbitals also have small contribution to the magnetic property of $Fe_3Se_4$. The difference of the spin-up band and spin-down band of Fe-3d orbitals show that they carry very large magnetic moment in $Fe_3Se_4$. Further, the spin polarization is negative near Fermi because of the electronic DOS of spin-down electrons is larger than that of spin-up electrons.

Furthermore, we have identified the orbital ordering (OO) state in $Fe_3Se_4$ as shown in projected DOS in Figure 3 with five electrons in 3d states of Fe ions in the 3D-coordinates (*xyz*) with the *x* and *y* axes pointing to the crystal [$\bar{1}01$] directions and *z* axis directed to the crystal *c*-axis. In the minority spin channel, four Fe ions (Fe-1, Fe-2, Fe-3 and Fe-4) show $t_{2g}$ bands right below the Fermi level ($E_F$) down to ~1.6 eV with reduced DOS at the $E_F$. All Fe ions in the unit cell of bulk $Fe_3Se_4$ can be categorized into two groups according to the orbital characters: spin-up/down $e_g$ bands fully unoccupied and spin-down $t_{2g}$ bands partially occupied, respectively. Bands of Fe-1, Fe-2, Fe-3 and Fe-4 [Figure 3(A, B, C and D)] are of predominate $d_{x^2-y^2}$ and $d_{z^2}$ characters, while bands of Fe-5 and Fe-6 [Figure 3(E) and 3(F)] are of mainly $d_{xy}$, $d_{z^2}$ and $d_{x^2-y^2}$ orbitals.

These results demonstrate that the configuration of ↓-spin $3dt_{2g}$ OO states in the Fe sub-lattice, where a ↓-spin electron of each Fe has occupied a mixed $t_{2g}$ state. The mixing of orbitals is a combination of two of the $t_{2g}$ states with the third empty $t_{2g}$ states[37]. In addition, it is found that the ↓-spin electrons of Fe from $3dt_{2g}$ OO state occupying, respectively, the canted $d_{xy}$ and $d_{yz}$ orbitals of Fe-1 to Fe-4 and $d_{yz}$ of Fe-5 and Fe-6 sublattice, as they contribute more at the Fermi level. The OO pattern is clearly seen in the distribution of charge density on each $Fe^{3+}$ with the distribution on Fe-1, Fe-2, Fe-3 and Fe-4 belonging to one state, while with Fe-5 and Fe-6 belonging to another state.

**Symmetry analysis**

According to the ionic model[37], we have acquired a metallic ground state in consistence. It is also shown that the tilting $3dt_{2g}$ OO states on the Fe sub-lattice is strongly related to the John-Teller (JT) distortions[37]. The main contributions in electronic structure were utilized for Fe ions to investigate the correlation effects in 3d-electrons. The OO states are generally found in $3de_g$ manganite frameworks where the helpful Jahn-Teller (JT) mutilations are huge because of the solid hybridization between the $e_g$ and O-2p electrons[38]. The $t_{2g}$ OO states with higher degeneracy and moderately weaker JT distortion is additionally found in confined 3d frameworks magnetite[39]. In the ionic model, the five 3d electrons of the $Fe^{3+}$ ion possess the $t_{2g}$ triplet degenerate and leave the higher $e_g$ doublet degenerate vacant, under the octahedral crystal field (see Figure S4 in ESI). As per Hund's rule, $Fe^{3+}$ is in the high spin states with the spin arrangement of ($t_{2g}^3 \uparrow t_{2g}^2 \downarrow$ ), giving rise to a magnetic moment of 4.25 $\mu_B$/unit cell: and it indicates metallic ground state with the majority $\uparrow$-spin at the Fermi level accountable for the conductivity. Assessing three 3d electrons, the spin up $t_{2g}$ states is moderately confined with a suppressed bandwidth and energy of lower band, while the $\downarrow$-spin $t_{2g}$ band is pushed upwards marginally [see ESI in Figure S1]. Besides, the $\uparrow$-spin $t_{2g}$ band is thus completely occupied and at $E_F$ there is no band gap, giving rise to a metallic ground state in consistence with the valence setup ($t_{2g}^3 \uparrow t_{2g}^2 \downarrow$ ) of Fe from ionic model as displayed in Figure S3 in ESI.

The tilting of $FeSe_6$ octahedral site (Fig. 1 (A)) is to accommodate two sorts of distortion of the Se lattice: the two Se ions at the tip of the octahedron deform along the *z*-axis (Fe-Se top bond length= 2.44 Å), while two ions of the coplanar Se move upwards and two ions move downwards along the *z*-axis. The JT distortions with Fe-Se bond length (2.46 Å, between Fe-1 and Se) along the *y*-axis and extended bond length (2.63 Å) along the *x*-axis (Figure 1) split the triply degenerate $t_{2g}$ orbitals into the lower $d_{xz}$ and higher $d_{yz}$ and $d_{xy}$ orbitals. The JT distortions could bring down the Coulomb interaction between the ions of Se and the Fe $d_{xz}$ states, which is a combination of $d_{xy}$ and $d_{yz}$ states as appeared in Figure S4 in ESI. Moreover, the $d_{xz}$ states with intermediate directions between nearest Se anions would additionally stabilize the lattice distortion.

It is well known that, if we include the spin-orbit coupling (SOC) interaction then the observed orbital-ordering in $SrRuO_3$ would be destroyed; otherwise if we do not include SOC interaction then the polyhedral crystal field makes the cubic harmonics (breaking the inversion symmetry)[40] a nature basis set resulting in lower $t_{2g}$ and higher $e_g$ bands[37]. The Se octahedral site also break parity in each site as Figure 1 shows that Se-5 is above the ladder's plane, but the next Se-7 is below, and the distances of Se-5 and Se-7 to the iron ladder plane should be the same in magnitude and opposite in sign ("antisymmetric"). However, the OO introduces a fundamental modification in the symmetry. Now the blocks made of four Fe ($\uparrow$) [or four Fe ($\downarrow$)] are no longer identical to pair of two Fe ($\uparrow$) and two Fe ($\downarrow$). Then, the heights of Se-5 and Se-7 do not require to be antisymmetric anymore; their distances to the *xz* plane can become different. A similar mechanism works for the edge Se's, e.g., Se -1 and Se-5. As a result, the atomic positions of Se break the space inversion symmetry, creating a local Ferroelectric (FE) Polarization (P) pointing perpendicular to the iron *yz* plane (almost along the *x*- axis).

To clarify more of the features of OO states, we have calculated charge density distribution (0.66e/Å$^3$) and depicted in Figure S5 in ESI which corresponds to the ↓-spin $t_{2g}$ bands below the Fermi level (-1.80-0.0 eV) [see ESI in Figure S1]. It has been observed that electron densities corresponding to ↑-spin electrons are totally different for each atom present in the zigzag whereas when we compare both the spins, electron densities are partially different from each other. This suggests that in case of $Fe_3Se_4$, spin down electrons present at unsaturated edge Fe-atoms are responsible for conduction as also confirmed through corresponding band structure and DOS. The charge contour of real space distribution of spin dependent electron densities suggests that the possibility of metallic spin polarization in $Fe_3Se_4$ structures is totally attributed to the localization of unpaired electrons at unsaturated Fe atoms which are situated at faces of the unit cell in Figure S5(A) and S5(B) in ESI.

The yellow spheres show that the Se anions attract more electrons while, the Fe cations lose more 3d electrons in $Fe_3Se_4$, shown by bright magenta lobes pointing along the Fe directions. By contrast, the density difference is weak but also exists in the Se ladder plane. There is a dark yellow sphere introduced as a Se atom, with negative value: this suggests that the outmost electrons of Fe are more extended (delocalized), also supporting the covalent scenario for $Fe_3Se_4$.

**Magnetic and ferroelectric properties**

The calculated total magnetic moment per unit cell of monoclinic $Fe_3Se_4$ with 14 atoms per unit cell (6-Fe atom and 8-Se atom) is 4.25 $\mu_B$. This value is consistent with the previous DFT calculations (4.34 $\mu_B$/unit cell)[35], but significantly higher than the reported experimental value for $Fe_3Se_4$ nanostructures (2.2 $\mu_B$/unit cell)[35]. This was attributed to the spin fluctuation and the long-range ordering as measured by experiments, which normally would be smaller than the calculated value at 0 K.

We have calculated spontaneous polarization (Ps) along each of the directions *x*, *y* and *z*, for the ferroelectric phase applying the Berry phase approach for bulk $Fe_3Se_4$. The direction of Ps in $Fe_3Se_4$ single crystal lies in the monoclinic *ac* plane and the magnitude of the segment of spontaneous polarization along the *a*-axis of the monoclinic unit cell. The spontaneous polarization (Ps) in $Fe_3Se_4$ has a value of ~44.60 μC/cm$^2$ and the direction of Ps vector makes an angle 91.15° with the major surface (normal to *c*-direction). The calculated values of three components of the spontaneous polarization vectors are $P_x$= 23.14 µC/cm², $P_y$= 6.20 µC/cm² and $P_z$=37.62 µC/cm², respectively. The magnitude of spontaneous polarization of $Fe_3Se_4$ is in very good agreement with previously reported experimental value[41]. Roy et al.,[42] provided the spontaneous polarization of ferroelectric bismuth titanate with a value of 42.83 µC/cm² which is fairly comparable to our calculated value for polarization. Another study by Ravindran et al.,[43] gave the spontaneous polarization of $BiFeO_3$ which is also within the reported agreement between theoretical and experimental values. Therefore, it is clear that the calculated values of spontaneous polarization using GGA fall in the range of 43-68 µC/cm².

**Dielectric and transport properties**

Furthermore, we have calculated the frequency dependent optical properties including dielectric function, absorption coefficient using DFT within the random phase approximation (RPA)[44]. The frequency dependent dielectric function can be written as $\varepsilon(\omega) = \varepsilon_1(\omega) + i\varepsilon_2(\omega)$. Where $\varepsilon_1(\omega)$ and $\varepsilon_2(\omega)$ are the real and imaginary parts of the complex dielectric function, respectively. $\varepsilon_2(\omega)$ is determined by summation over electronic states and $\varepsilon_1(\omega)$ is obtained using the Kramers–Kronig (KK) relationship[45].

The real and imaginary parts of the complex dielectric function $\varepsilon$ versus frequency for bulk $Fe_3Se_4$ are presented in Figure (4A and 4B). The real ($\varepsilon_1$) and imaginary ($\varepsilon_2$) part follow the general tendency of a metal which can be well explained by the classical Drude theory. In Figure 4A and 4B, with increasing in frequency, the $\varepsilon_1$ decreases while $\varepsilon_2$ initially increases and decreases above the $1.5 \times 10^{14}$ Hz up to $2.0 \times 10^{14}$ Hz region, as expected in a metal[46]. The variation of dielectric constant $|\varepsilon|$ and loss tangent is shown in Figure 4C and 4D. With increasing frequency, the $|\varepsilon|$ decreases and $\tan\delta$ initially increases at certain frequencies and then it is decreases in all the polarization direction (in-plane and out-of-plane). The loss tangent measures the loss-rate of power in an oscillatory dissipative system[47]. It is clearly seen that the variation $\tan\delta$ executes in the same trend as $\varepsilon_2$. Since $\varepsilon_2$ is lower than $\varepsilon_1$, then the energy loss of the materials is relatively low. This suggests that the material possesses good optical qualities due to lower energy losses and lower scattering of the incident radiation[47]. Our theoretical results of the dielectric function of bulk $Fe_3Se_4$ are in excellent agreement with previously reported experimental results[48] except $\varepsilon_2$. But according to other theoretical investigation our results is expected in a metal[46].

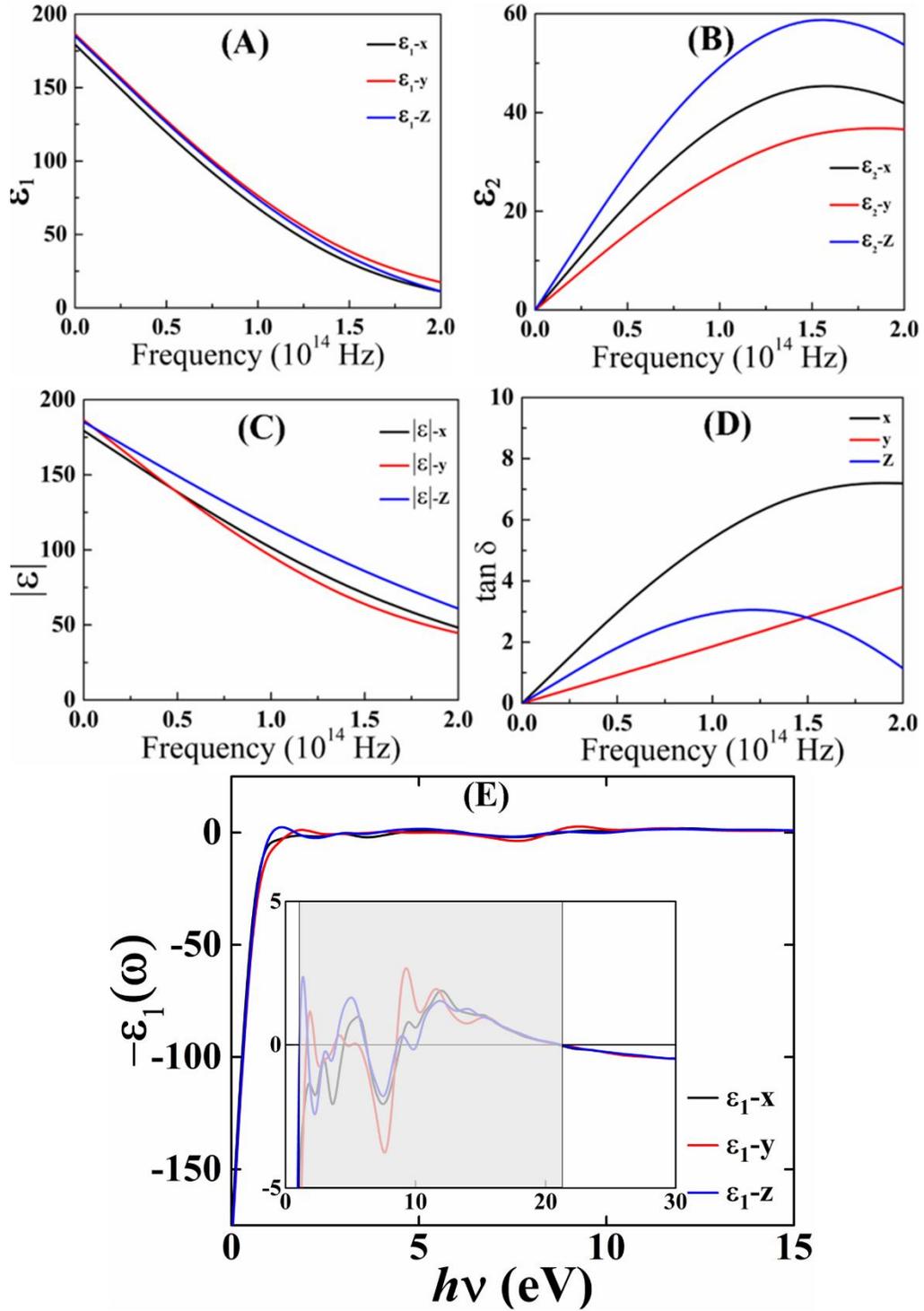

**Figure 4**. (A) Real and (B) imaginary part of the complex dielectric constant, (C) magnitude of the complex dielectric function and (D) loss tangent verses frequency at $10^{14}$ Hz at room temperature (300 K) of bulk $Fe_3Se_4$. (E) negative value of real part of complex dielectric function is a function of photon energy.

The real part of complex dielectric function $\varepsilon_1(\omega)$ is shown in Figure 4(E). In Figure 4(E), we plot the real parts of dielectric functions in all the polarisation direction for the materials. Character of this material exhibit metallic behavior, i.e. Drude peaks at low energy due to intraband contribution and the real part of $\varepsilon(\omega)$ crossing from negative value to positive value with increasing frequency and it's shows oscillatory behavior upto 15 eV. The strong anisotropy in the real part of complex dielectric function is also quite obvious. Remarkably, because of this anisotropy, the real part of complex dielectric function, $\varepsilon_1(\omega)$ in x, y and z-direction, changes sign at a frequency which is different from each other, leading to an extended frequency window in which the each components have different signs shown in shaded region with gray color in Figure 4(E). Such type of sign difference is the characteristic feature in real part of dielectric function is known as indefinite media[49]. Generally, indefinite media are mostly in artificially assembled structures which require complicated fabrication process and usually have high dissipation. This theoretical results suggest that crystalline solid $Fe_3Se_4$ in the bulk form would just be indefinite materials for a frequency range spanning the near infrared.

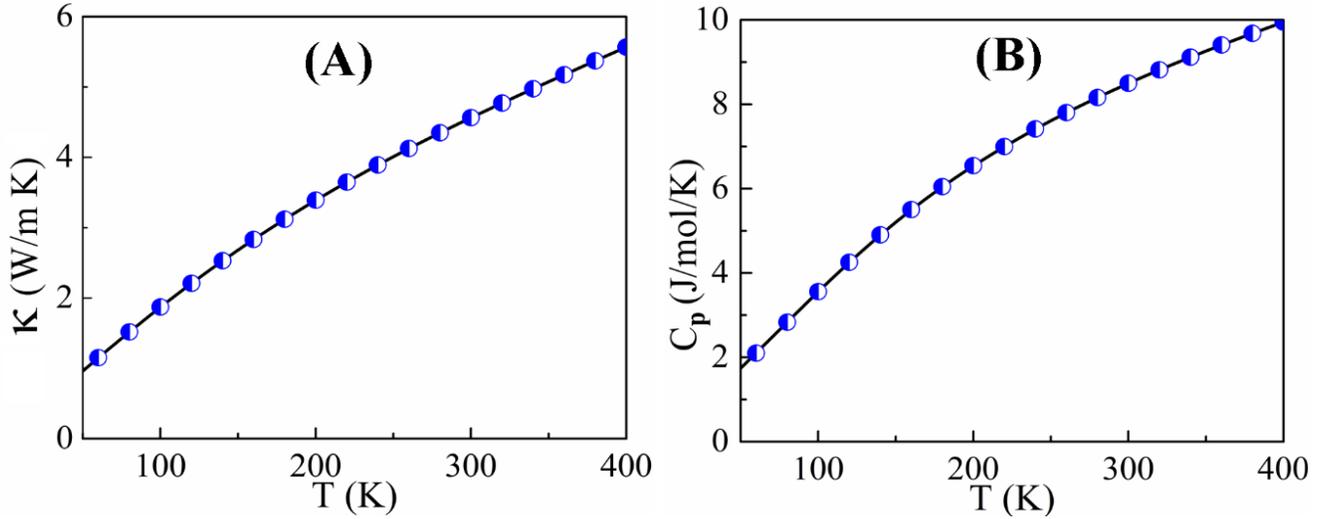

**Figure 5.** The variation of (A) thermal conductivity $\kappa$, (B) heat capacity $C_p$ as a function of temperature.

The thermal conductivity ($\kappa$) of a material originates from two fundamental sources: (i) electrons and holes transporting heat ($\kappa_e$) and (ii) phonons travelling through the lattice ($\kappa_l$). Most of the electronic term ($\kappa_e$) is straight forwardly associated with the electrical conductivity which can be well explained by Wiedemann–Franz hypothesis[50]. From Figure 5(A), the thermal conductivity of $Fe_3Se_4$ is almost linearly increased from 50 K to 400 K, and at room temperature the thermal conductivity is 4.56 W/ (m K). The thermal conductivity is dominated by the lattice contribution as it is about 1.0 W/m K at T=50 K and it increases to 5.57 W/mK at T=400 K which is higher than the values observed in most thermoelectric materials[50, 51]. Since the heat flow in a material is directly proportional to its thermal conductivity, the heat flow in bulk $Fe_3Se_4$ will be higher as well.

Specific heat estimation is one of the most reliable techniques for exploring temperature reliance of materials. The heat capacity $C_p$ of bulk $Fe_3Se_4$ increases with temperature as shown in Figure 5(B), demonstrating a similar behavior as the thermal conductivity. In addition, the variation of $C_p$ with temperature follows the well-known Debye form and shows excellent fit at lower temperatures. Our theoretical results are in good agreement with experimental results[48]. Furthermore, the estimated electric conductivity as a function of temperature of $Fe_3Se_4$ reveals nearly a linear behavior with a positive slope as plotted in Figure 6. This could be ascribed to the way that the electrical conductivity was evaluated based on DFT by assuming a constant scattering time $\tau(=10^{-14} s)$. In general, in case of metallic systems, the electrical conductivity decreases with respect to temperature owing to the occurance of more collisions among electrons and between electrons and phonons which shorten the mean free path of charge carriers. In our case, similar behavior are found in its metallic system, the electrical conductivity of $Fe_3Se_4$ continuously decreases with increasing temperature. This is likely due to the enhanced delocalization of hybrid orbitals.

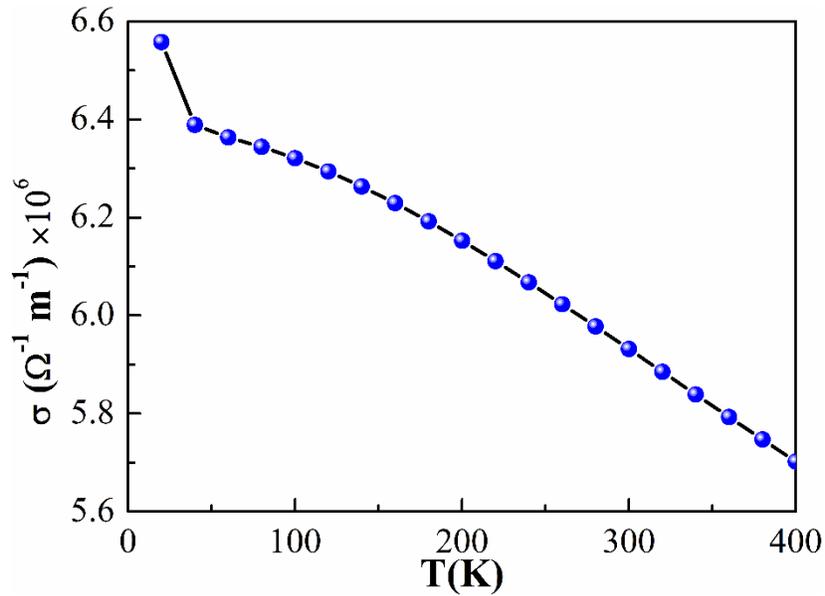

**Figure 6.** The variation of electronic conductivity $\sigma$ as a function of temperature.

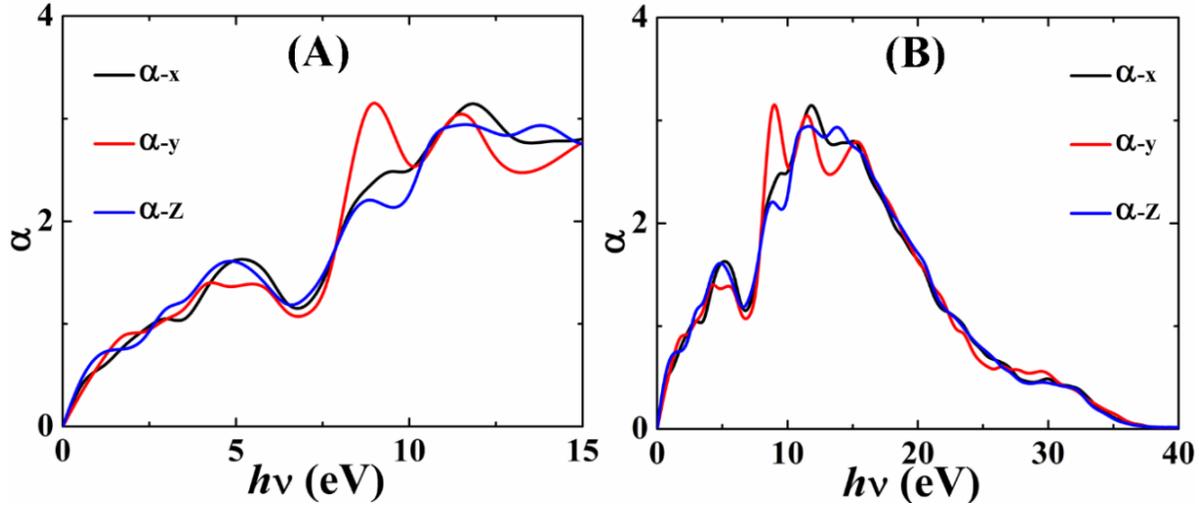

**Figure 7.** The absorption coefficient $\alpha$ in the unit of $10^5$/cm of bulk $Fe_3Se_4$ for all the polarization direction in the electric field. (A) The absorption coefficient up to 15 eV and (B) absorption coefficient up to 40 eV beyond 28 eV the absorption will be almost zero.

The absorption coefficient $\alpha$ in all the polarization directions of the electric field (depicted in Figure 7) show two main peaks in all polarizations. The first main peak occurs at energy around 5 eV that is related to π electron plasmonic peaks. And other peak occurs around 15 eV that is associated to π+σ electron plasmonic peaks. Moreover, at the energy range of 10-20 eV, the value of absorption for all cases is very high.

**Conclusions**

In conclusions, we investigated the novel multiferroic material shows ferroelectric, ferrimagnetism and electronic structure in monoclinic $Fe_3Se_4$ material. The nature of orbital ordering, and its close connection to the JT distortion is unwound, which is responsible for ferroelectric like instability. $Fe_3Se_4$ takes a ferromagnetic ordering with a total magnetic moment of 4.25 $\mu_B$ per unit cell (6 Fe atom and 8 Se atom). The spontaneous polarization (44.60 μC/cm$^2$) provides evidence of its ferroelectric behavior. Our study suggests that the monoclinic phase of $Fe_3Se_4$ material possesses the unusual dual behavior of metallicity and ferroelectricity. The stabilization of the ferroelectric structure in $Fe_3Se_4$ coexisting with metallic conductance is the consequence of a decoupling between the metallic electrons in the $t_{2g}$ electrons from the soft phonons which break the inversion symmetry. The development of multiferroic material are usefull applications in spintronics-related technologies for ultrahigh-density memory and quantum-computer devices is underway.

**Acknowledgements**

Helpful discussion with Drs. Pankaj Poddar and Mousumi Sen is acknowledged. S. K. G. acknowledges the use of high performance computing clusters at IUAC, New Delhi and YUVA, PARAM II, Pune to obtain the partial results presented in this paper. S. K. G and Y. A. S also thank the Science and Engineering Research Board (SERB), India for the financial support (grant numbers.: YSS/2015/001269 and EEQ/2016/000217, respectively). D. S. would like to thank University Grant Commission (UGC), New Delhi, India for the financial support.